\documentclass[11pt,a4paper]{article}
\usepackage[english]{babel}
\pdfoutput=1 
\usepackage{jinstpub} 
\usepackage{lineno}
\usepackage{siunitx}
\usepackage{subcaption}
\usepackage{breakurl}
\usepackage{ulem}
\usepackage{microtype}
\usepackage{makecell}
\usepackage{csquotes}
\usepackage{enumitem}
\usepackage{chemformula}
\usepackage{mathtools}
\usepackage[nameinlink]{cleveref}
\usepackage{svg}
\usepackage{doi}
\usepackage{xurl}
\DeclareSIUnit\sample{s}
\DeclareSIUnit\litre{l}
\DeclareSIUnit\liter{l}
\setitemize{topsep=0pt,parsep=0pt,partopsep=0pt}
\title{\boldmath Performance study of large-area glass resistive plate chambers with different spacer configurations}

\author{L. Mao$^a$,}
\author{F. Lagarde$^a$,}
\author[1]{J. Guo$^{a,}$\note{Corresponding authors}}
\author{,  X. Wang$^a$,}
\author{J. Li$^a$,}
\author{Q. Shen$^a$,}
\author{J. Zhu$^a$,}
\author[1]{H. Yang$^{a,b,}$}

\affiliation{$^a$Institute of Nuclear and Particle Physics, School of Physics and Astronomy, Shanghai Jiao Tong University\\
Key Laboratory for Particle Physics, Astrophysics and Cosmology (Ministry of Education)\\ 
Shanghai Key Laboratory  for Particle Physics and Cosmology\\
800 Dongchuan Road, Shanghai, 200240, P.R. China}
\affiliation{$^b$Tsung-Dao Lee Institute, Shanghai Jiao Tong University, Shanghai, 200240, P.R. China}

\emailAdd{jun.guo@sjtu.edu.cn, haijun.yang@sjtu.edu.cn}

\abstract{Optimization of spacer and gas distribution inside the glass resistive plate chamber (RPC) is reported. Simulation studies demonstrate improvements on the gas flow velocity homogeneity and lower vorticity inside the gas chamber. The optimized spacer configuration (\num{76} spacers) decreases the number of spacers by 24\% compared to the original design (\num{100} spacers), thus helps significantly reduce the non-active or low-efficiency area caused by spacers while maintaining similar deformation uniformity of the electrodes. Large area glass RPCs with \num{1}$\times$\SI{1}{\square\meter} size using two types of spacer configurations are constructed and tested with cosmic muons events. The muon detection efficiencies for RPCs are greater than 95\%.}

\keywords{Gaseous detector, Resistive plate chamber, Calorimeter, Muon spectrometer}

\begin{document}
\maketitle
\flushbottom

\section{Introduction}

Since the invention of the Resistive Plate Chamber (RPC) in 1980s, it has been successfully used in many particle physics experiments such as \textsf{ATLAS} \cite{Aad:2753039}, \textsf{CMS} \cite{Shah_2020}, \textsf{ALICE} \cite{boss2012performance} at \textsf{CERN}, astrophysics experiments such as  \textsf{ARGO-YBJ} \cite{BARTOLI201547,BACCI2003110} at Yangbajing, neutrino physics experiments such as \textsf{Daya Bay} \cite{Ning_2013}.  RPCs can be used in many other applications such as time of flight positron emission tomography (PET/CT), and volcano tomography (\textsf{TOMUVOL}) \cite{Crloganu2012TowardsAM} etc...

The interests for these gaseous detectors can be explained by their excellent features: good spatial ($\mathcal{O}$(\SI{1}{\centi\meter}) \cite{CATTANI2012S6,ARNALDI200251}) and time ($\mathcal{O}$(\SI{1}{\nano\second}) \cite{AIELLI2004193,ARNALDI2000462,WILLIAMS2002183,BERTOLIN2009631,AIELLI200692,ZHANG2007278,ABBRESCIA2003137,AIELLI2013115}) resolution, robustness, easy to build and relative low cost, which make the RPC well suited for particle detection and triggering.  In addition to this "traditional" application, the RPC can also be used for high performance calorimetry based on the Particle Flow Algorithm (PFA)\cite{PFA}.

RPCs are expected to play an important role in new particle detectors such as the Semi-Digital Hadronic Calorimeter (\textsf{SDHCAL}\cite{SDHCAL}) at the  International Linear Collider (\textsf{ILC}) \cite{baer2013international} and the Circular Electron Positron Collider (\textsf{CEPC}) \cite{thecepcstudygroup2018cepc,Yu:2019bm}. 

The design of the RPC should in principle minimize the non-active area and maximize detection efficiency.  It is also important to maintain the relative simplicity of construction process and keep the cost advantage. We present the results obtained when optimizing the position and number of spacers to maintain the flatness of the electrodes, which will also lead to a slight improvement of the gas flow uniformity inside the chamber.

\newpage
We use gas flow simulations to optimize the design of the glass RPC and present a new shifted-spacer design which leads to an improved gas flow uniformity, a decrease in the number of spacers by around 24\% while maintaining the flatness of the chambers ($\sim\num{0.5}\%$) at the level of the baseline design. In this paper, the results of the prototype chambers built based on the new design are compared to the one built based on the baseline design.

\section{RPC Design and Gas flow Simulations}

The operation of the RPC requires continuous gas flow at a reasonable rate to avoid accumulation of polluted gas inside the chamber caused by the creation of radicals and the interaction of the radicals produced in the gas with the materials of the chamber \cite{AIELLI2006143}. The uniformity of the gas flow in the chamber and the deformation of the electrode plates are critical to the performance of the RPC. Detailed simulations help optimizing the design of the RPC. We use \textsf{COMSOL Multiphysics\textsuperscript{®}} \cite{COMSOL} version \num{5.4} as a platform for simulation which implements a finite element method, through which the behaviour of the flow field, electrode deformation and electric field are simulated.

\subsection{Design and geometry for the RPC}

\begin{figure}[ht!]
  \centering
  \includegraphics[width=0.99\textwidth]{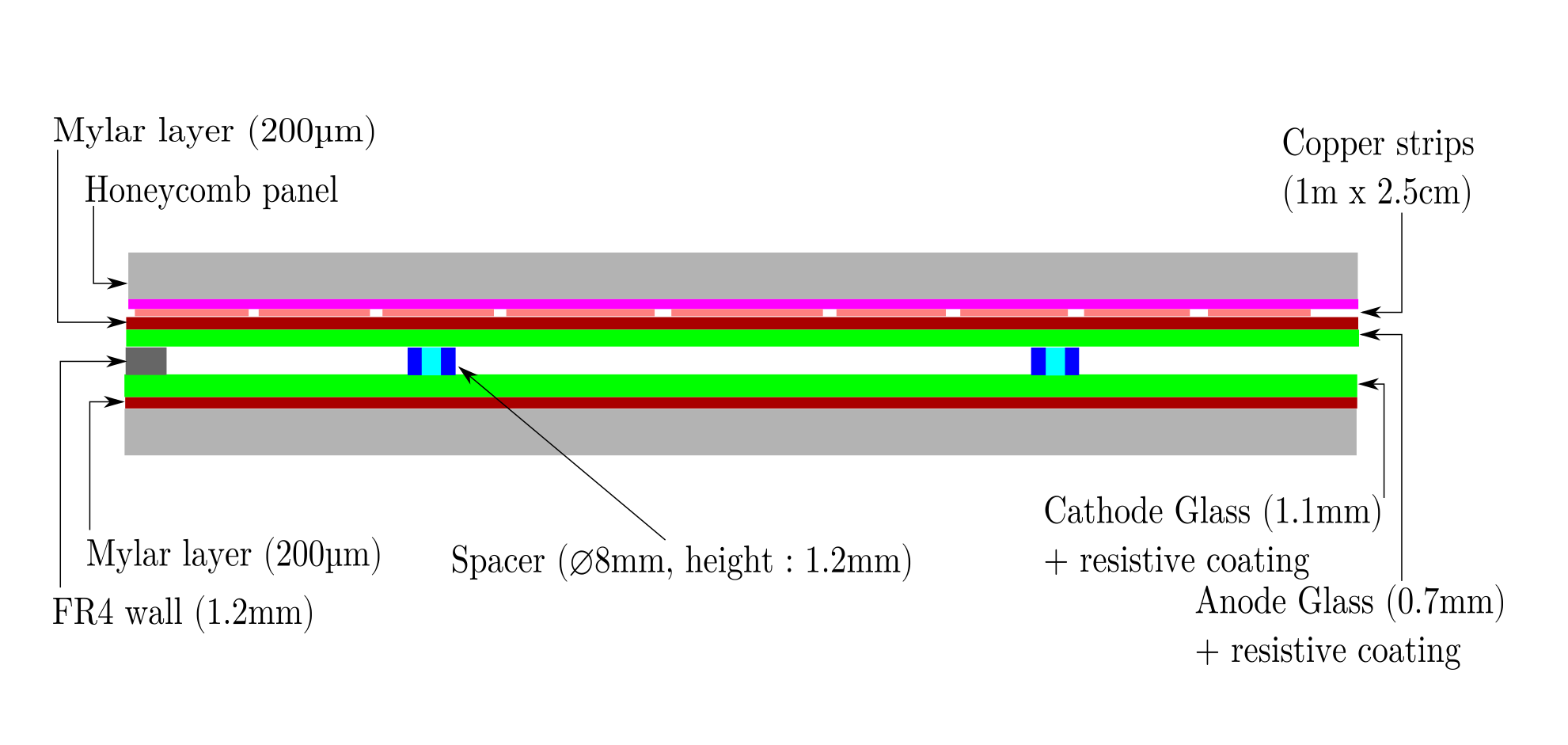}
  \caption{Schematic cross section of  the glass RPC (not to scale).}
  \label{fig:RPC}
\end{figure}

Glass RPCs are constructed using two pieces of glass (\num{1}$\times$\SI{1}{\square\meter}) with thickness of \SI{0.7}{\milli\meter} (anode) and \SI{1.1}{\milli\meter} (cathode). The gap between the two electrodes is \SI{1.2}{\milli\meter} wide, sustained by cylindrical nylon spacers with a diameter of \SI{8}{\milli\meter}. The gas gap is sealed by the surrounding FR4 walls and silicone-based glue. The gas inlet and outlet are located at the two opposite corners along the same side of the chamber, perforated with \num{2} pipes for the inlet and \num{4} pipes for the outlet to avoid over-pressure. Two gas guides made of  \textsf{PEEK} and perforated with \num{6} holes are placed  at \SI{2}{\centi\meter} from the walls along the inlet side and outlet side to uniformly distribute the gas into the chamber. The schematic cross section of the RPC is shown in Figure~\ref{fig:RPC}.

\begin{figure}[ht!]
  \centering
  \begin{subfigure}{0.45\textwidth}
    \centering
	\includegraphics[width=1.02\textwidth]{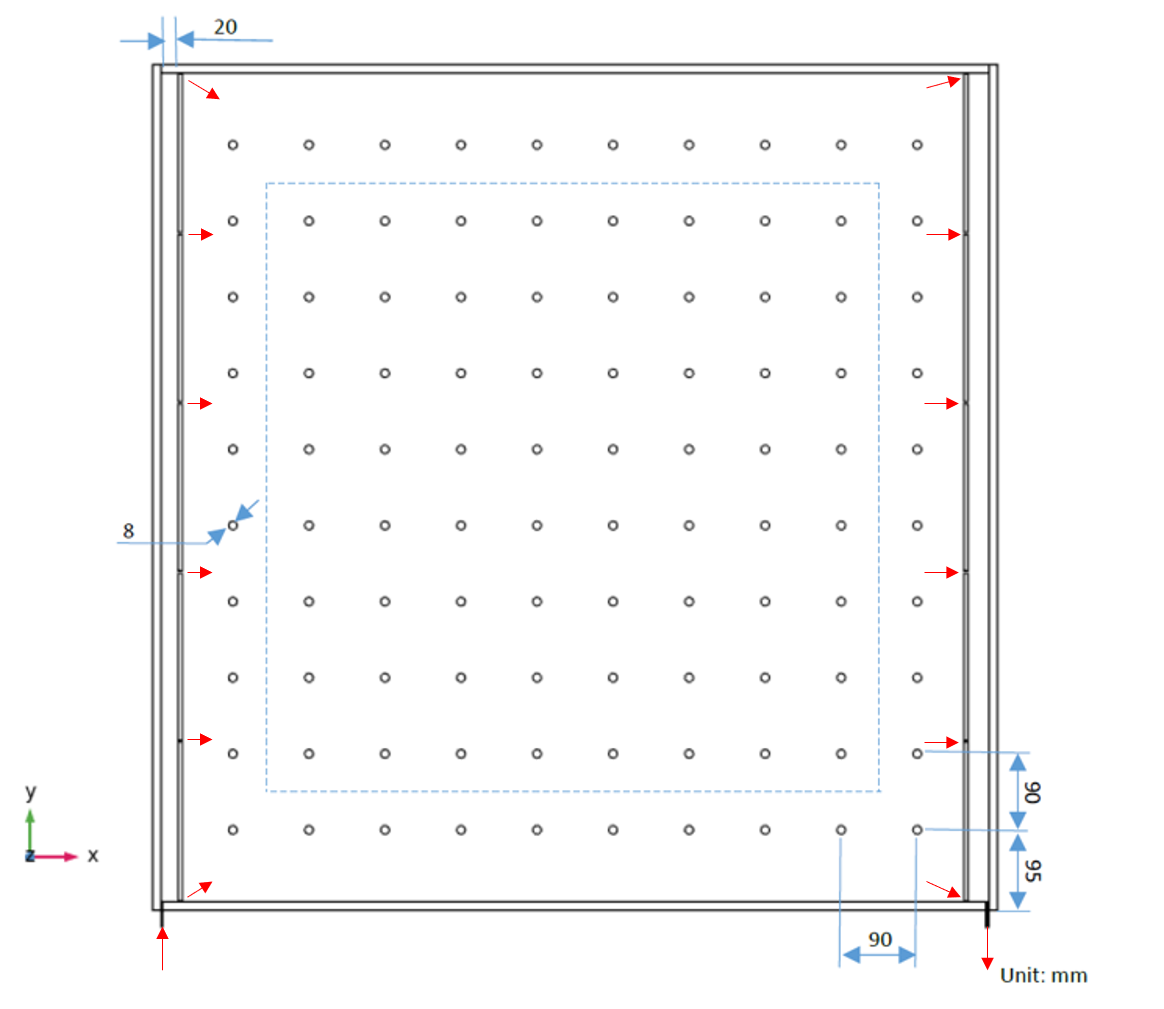}
	\caption{\label{fig:newRPCa}"Reference spacers" RPC.}
  \end{subfigure}%
  \begin{subfigure}{0.45\textwidth}
    \centering
	\includegraphics[width=1.02\textwidth]{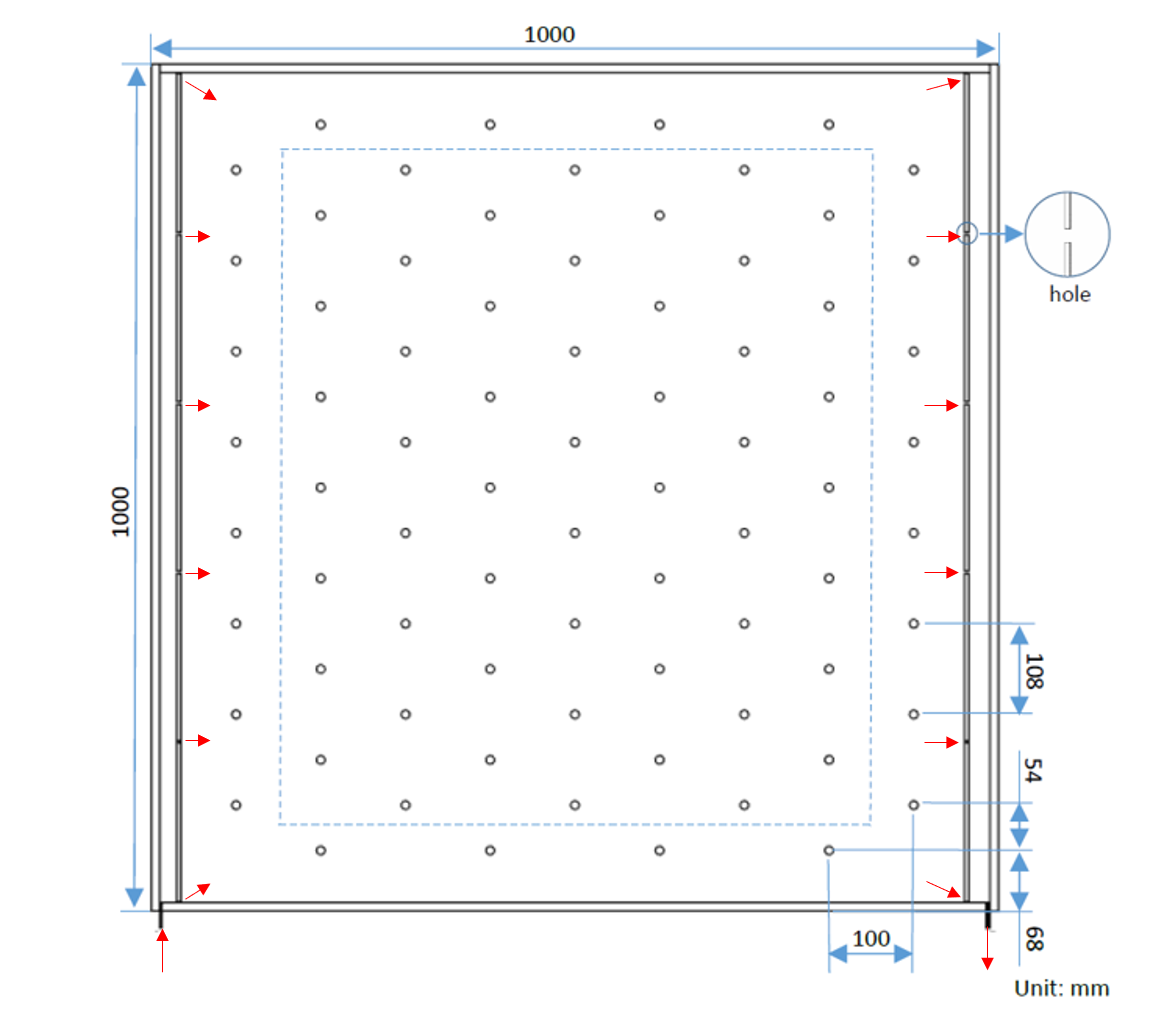}
	\caption{\label{fig:newRPCb}"Shifted spacers" RPC.}
  \end{subfigure}
  \caption{Drawing of \num{1}$\times$\SI{1}{\square\meter} RPCs with two different types of spacer configurations. The central part of the chamber marked by the dashed lines is used for gas velocity and vorticity comparison.}
  \label{fig:newRPC}
\end{figure}

\newpage
Spacers play a key role in keeping the thickness uniformity of the gas gap, but they are an obstacle for smooth gas diffusion. Two types of spacer distributions are presented. One is  a "Reference spacers" RPC with standard aligned spacers (100 spacers). There are a grid of $10 \times 10$ spacers, as shown in Figure~\ref{fig:newRPCa}. The second one is a "Shifted spacers" RPC with \num{4} spacers on the odd line and \num{5} spacers on the even lines, which has \num{76} spacers in total, as shown in Figure~\ref{fig:newRPCb}.

\subsection{Simulation of the RPC gas flow}

A gas flow simulation is performed to simulate the flow field and it follows the Navier-Stokes equations:

\begin{equation}
 \rho(\boldsymbol{u}\cdot\nabla)\boldsymbol{u} = \nabla\cdot[-\overset{\rightharpoonup\!\!\!\!\rightharpoonup}{p}\boldsymbol{I} + \mu(\nabla\boldsymbol{u}+(\nabla\boldsymbol{u})^{T})] + \boldsymbol{F} ; 
\end{equation}

\begin{equation}
 \rho\nabla\cdot\boldsymbol{u} = 0 ,
\end{equation}
where $\boldsymbol{u}$ is the velocity of the flow field, $\rho$ is the density of the gas, $\overset{\rightharpoonup\!\!\!\!\rightharpoonup}{p}$ is the stress tensor, $\mu$ is the dynamic viscosity of the gas, $\boldsymbol{F}$ is the external force and $\boldsymbol{I}$ is the identity matrix. The boundary conditions are fixed by the walls. The input flow rate of the gas is \SI{1}liter per hour, corresponding to a mean velocity of injected gas of \SI{0.5526}{\meter\per\second}.  The external pressure is fixed to zero as a reference value.

The electric field is given by the Maxwell equations:

\begin{equation}
 \nabla\cdot\boldsymbol{D} = \rho_e ;
\end{equation}
\begin{equation}
 \boldsymbol{E} = -\nabla V ,
\end{equation}

where $\boldsymbol{D}$ is the electric displacement, $\rho_e$ is the electric charge density, $\boldsymbol{E}$ is the electric field  and $V$ is the potential.

The deformation of the electrodes are obtained using a structural mechanics simulation that solved the following equations:
\vspace*{-0.1cm}
\begin{equation}
  \nabla\cdot(\overset{\rightharpoonup\!\!\!\!\rightharpoonup}{F}\overset{\rightharpoonup\!\!\!\!\rightharpoonup}{S})^{T} + \overset{\rightharpoonup\!\!\!\!\rightharpoonup}{F}\nu  = 0 ;
\end{equation}
\vspace*{-0.5cm}
\begin{equation}
 \overset{\rightharpoonup\!\!\!\!\rightharpoonup}{F} = \boldsymbol{I} + \nabla\boldsymbol{d} ,
\end{equation}
\vspace*{-0.1cm}
where $\overset{\rightharpoonup\!\!\!\!\rightharpoonup}{S}$ is the stress tensor,  $\nu$ is the Poisson ratio of the material,  $\overset{\rightharpoonup\!\!\!\!\rightharpoonup}{F}$ is the deformation gradient tensor and $\boldsymbol{d}$ is the displacement vector.  In this simulation only the chamber has been studied (no cassette, metallic panels parceling the RPCs which act as electromagnetic shielding) and only the pressure of the gas and the electric force are considered. 
 
To couple the flow field to the stress and deformation tensors, the pressure is extracted from the gas flow simulation and used to calculate the force exerted by the gas to the glass.

The coupling between electric field and structural mechanics can be derived from the following thermodynamic potential called electric enthalpy \cite{MEMS} :

\vspace*{-0.3cm}
\begin{equation}
 H_{eme} = W_s(\overset{\rightharpoonup\!\!\!\!\rightharpoonup}{C}) - \frac{1}{2}\varepsilon_{0}\varepsilon_{r} J \overset{\rightharpoonup\!\!\!\!\rightharpoonup}{C}^{-1} : \boldsymbol{E}\otimes\boldsymbol{E} ,
\end{equation}
\vspace*{-0.3cm}

where $\overset{\rightharpoonup\!\!\!\!\rightharpoonup}{C}$ = $(\overset{\rightharpoonup\!\!\!\!\rightharpoonup}{F})^{T}$$\overset{\rightharpoonup\!\!\!\!\rightharpoonup}{F}$ is so called right Cauchy-Green deformation tensor generated from the deformation gradient tensor $F$ mentioned above, $J$ = det($\overset{\rightharpoonup\!\!\!\!\rightharpoonup}{F}$) and $W_s$($\overset{\rightharpoonup\!\!\!\!\rightharpoonup}{C}$) is the mechanical energy function which depends on the solid model that has been used. The product $:$ is the double dot product of two tensors and $\otimes$ is the tensor product of two vector spaces. Then we can obtain the stress tensor for deformation and the electric displacement by:

\vspace*{-0.2cm}
\begin{equation}
 \overset{\rightharpoonup\!\!\!\!\rightharpoonup}{S} = 2\frac{\partial H_{eme}}{\partial \overset{\rightharpoonup\!\!\!\!\rightharpoonup}{C}} ;
\end{equation}

\vspace*{-0.2cm}
\begin{equation}
 \boldsymbol{D} = -\frac{\partial H_{eme}}{\partial \boldsymbol{E}} 
\end{equation}
\vspace*{-0.8cm}

\subsection{Optimization of the RPC design}
\begin{figure}[htp!]
  \centering
  \begin{subfigure}{.45\textwidth}
    \centering
	\includegraphics[width=1\textwidth]{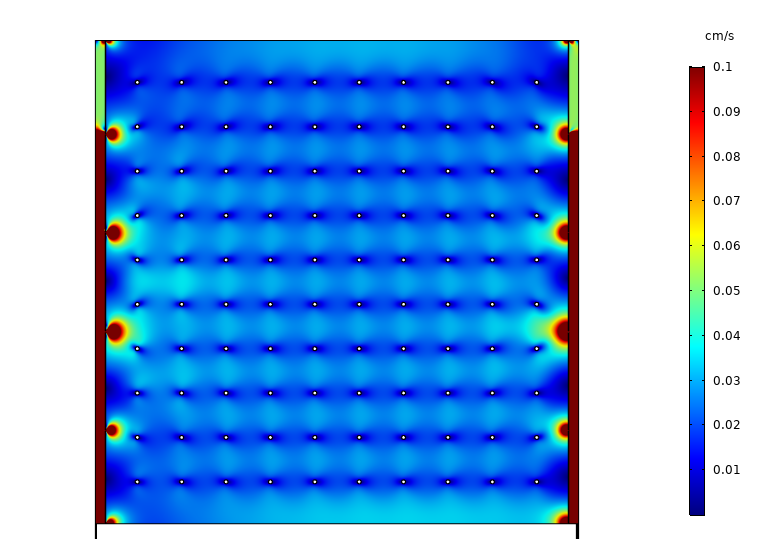}
	\caption{"Reference spacers" RPC.}
  \end{subfigure}
  \begin{subfigure}{.45\textwidth}
    \centering
	\includegraphics[width=1\textwidth]{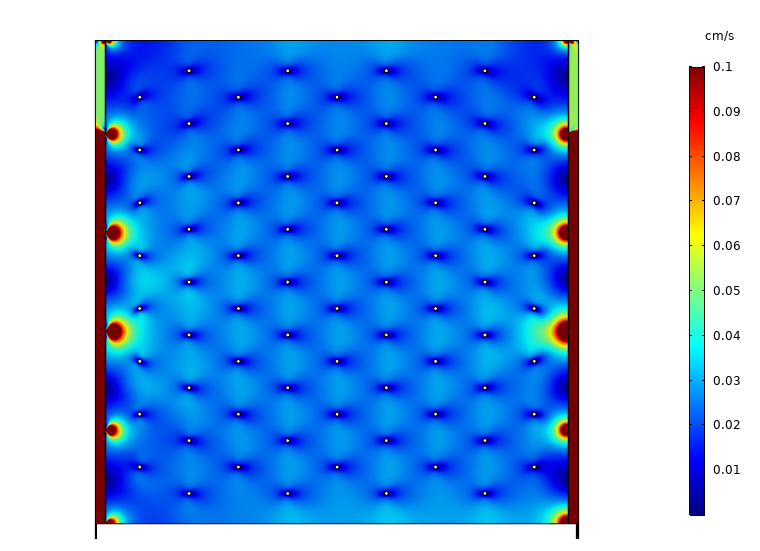}
	\caption{"Shifted spacers" RPC.}
  \end{subfigure}
  \caption{Velocity distribution inside the gas chamber.}
  \label{fig:velocity}
\end{figure}

\begin{figure}[thp!]
  \vspace*{-0.2cm}
  \centering
  \begin{subfigure}{0.45\textwidth}
    \centering
    \includegraphics[width=1\textwidth]{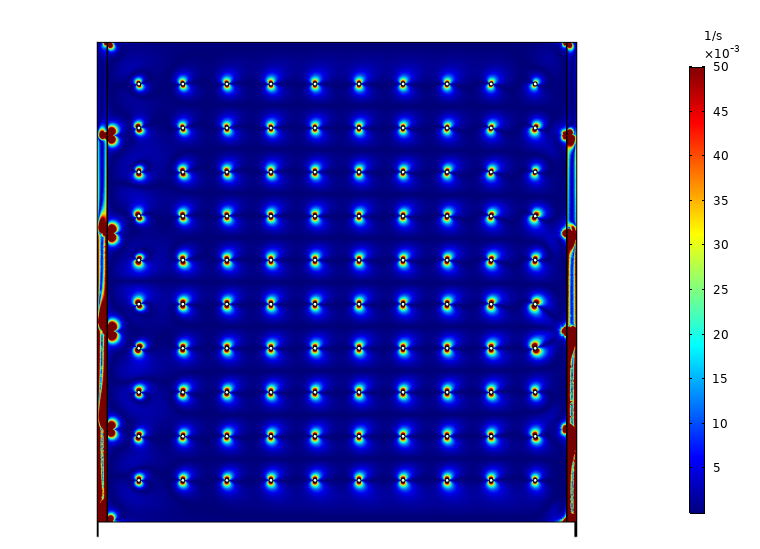}
    \caption{"Reference spacers" RPC.}
  \end{subfigure}
  \begin{subfigure}{0.45\textwidth}
    \centering
	\includegraphics[width=1\textwidth]{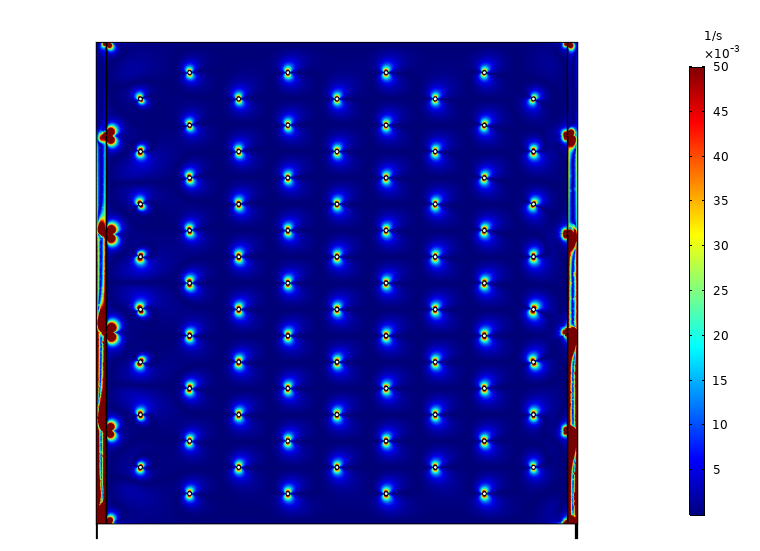}
	\caption{"Shifted spacers" RPC.}
  \end{subfigure}
  \caption{Vorticity distribution inside the gas gap.}
  \label{fig:vorticity}
  \vspace*{-0.6cm}
\end{figure}

\vspace*{-0.1cm}
The gas flow simulation uses an input flow rate of \SI{1} liter per hour, corresponding to a mean velocity of injected gas \SI{0.5526}{\meter\per\second}, which is equivalent to replace $\sim$\num{80}\% of the gas chamber volume per hour. The velocity and vorticity distributions inside the chamber are shown in Figure~\ref{fig:velocity} and Figure~\ref{fig:vorticity}, respectively.
\vspace*{-0.3cm}

In order to avoid the border effect, we use the central part of the chamber (surrounded by the dashed line in Figure~\ref{fig:newRPC}) to calculate distributions of gas velocity and vorticity for comparison. The mean velocity of gas inside the "Reference spacers" chamber and the "Shifted spacers" chamber are \SI{0.24}{\milli\meter\per\second} and \SI{0.24}{\milli\meter\per\second} respectively. The corresponding ratios (RMS/mean) of root-mean-square to mean velocity are \SI{20.3}\% and \SI{17.5}\% respectively, which indicate that the "Shifted spacers" RPC provides a more uniform distribution of gas flow than the "Reference spacers" RPC.

As for vorticity (defined by $\nabla\times\boldsymbol{u}$), there exist high vorticity regions around the gas entry and exit, in the gas tunnel and around the spacers as shown in  Figure~\ref{fig:vorticity}. To better estimate the vorticity near the spacers, we choose a region around the spacers (with a radius of \SI{12}{\milli\meter} with respect to the spacer center); The mean vorticities around spacers of the "Reference spacers" chamber and the "Shifted spacers" chambers are \SI{0.0199}{\per\second} and \SI{0.0196}{\per\second} respectively.
The mean vorticities for the remaining volume in the central part of chambers are \SI{0.0022}{\per\second} and \SI{0.0018}{\per\second} respectively. It shows that shifting the spacers and reducing the number of spacers can reduce the vorticity inside the gas chamber. The RMS value for each case is shown in table \ref{tab:simulation}.

\begin{figure}[htp!]
 \vspace*{-0.4cm}
  \centering
  \begin{subfigure}{.47\textwidth}
    \centering
	\includegraphics[width=1\textwidth]{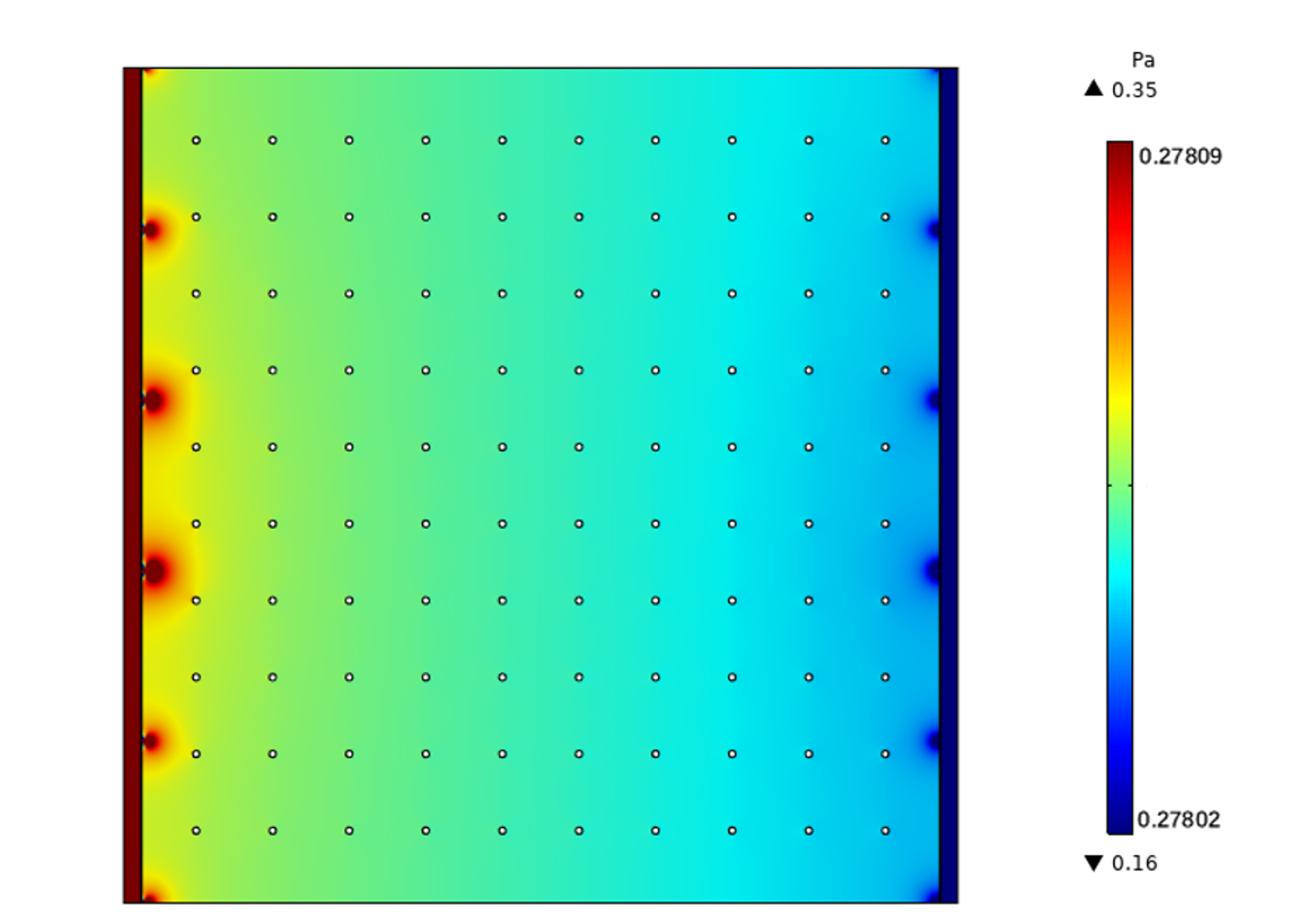}
	\caption{"Reference spacers" RPC.}
  \end{subfigure}
  \begin{subfigure}{.47\textwidth}
    \centering
	\includegraphics[width=1\textwidth]{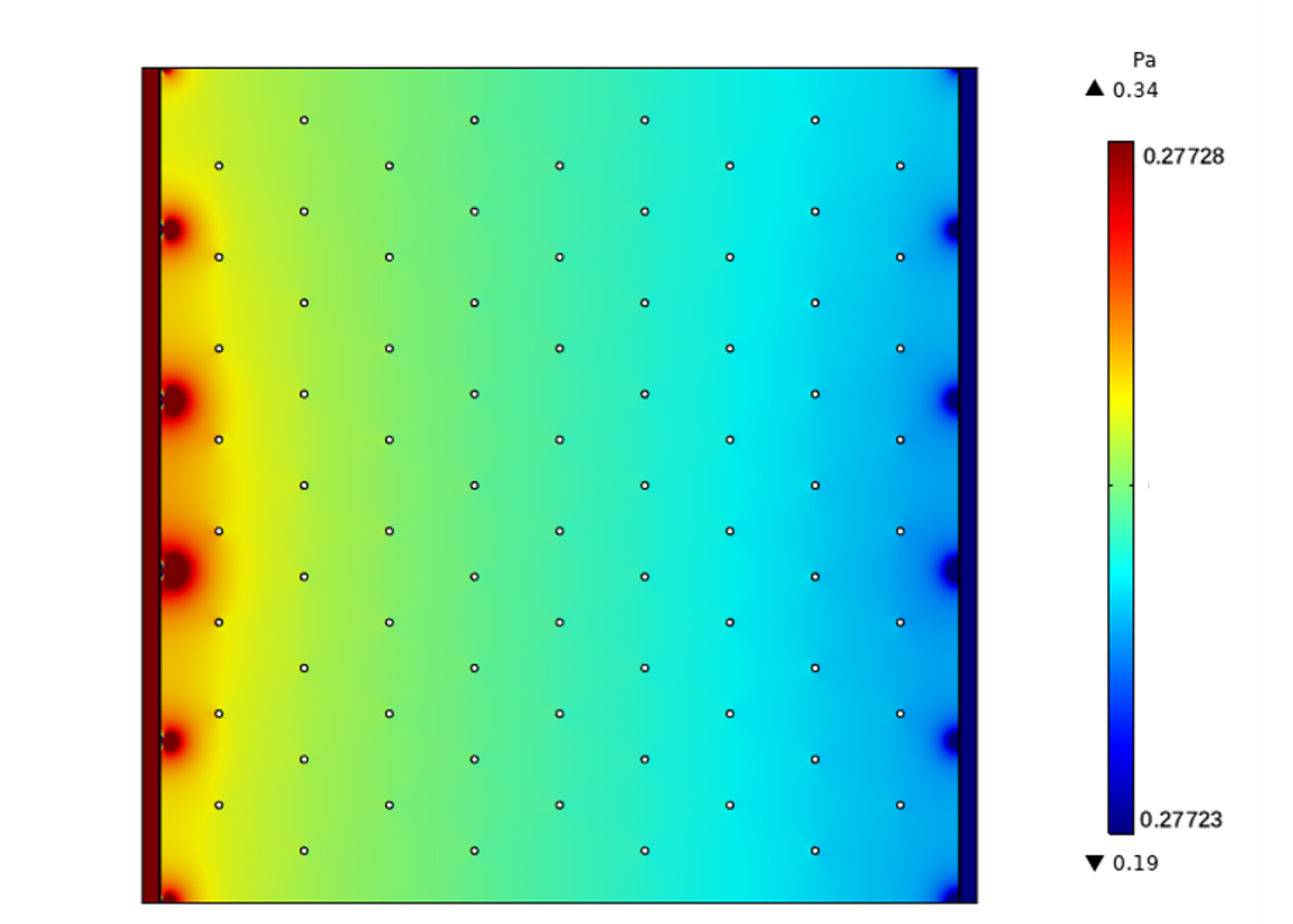}
	\caption{"Shifted spacers" RPC.}
  \end{subfigure}
  \caption{Pressure distribution inside the gas chamber.}
  \label{fig:pressure}
\end{figure}

The pressure distributions inside the gas chambers are shown in Figure~\ref{fig:pressure}. The lower limits and upper limits for each color scale in the graphs are [0.27802, 0.27809](Pa) and [0.27723, 0.27728](Pa), respectively. The distribution maps demonstrate that the pressure drops inside the chamber from inlet side to outlet side but with only a tiny value.

\begin{figure*}[ht!]
  \centering
  \begin{subfigure}{0.45\textwidth}
    \centering
	\includegraphics[width=1\textwidth]{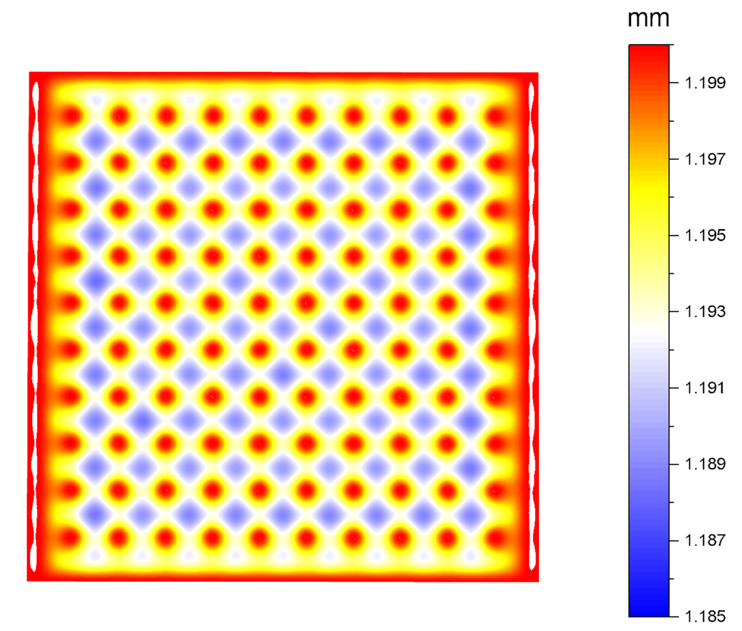}
	\caption{"Reference spacers" RPC.}
  \end{subfigure}
  \begin{subfigure}{0.45\textwidth}
    \centering
	\includegraphics[width=1\textwidth]{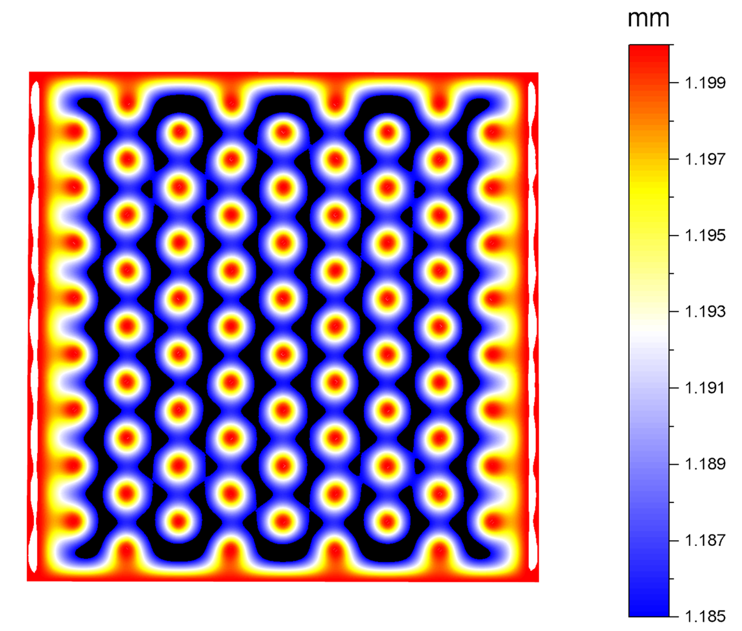}
	\caption{"Shifted spacers" RPC.}
  \end{subfigure}
  \caption{Deformation map of the thickness of the gas chamber.}
  \label{fig:deformation}
\end{figure*}

\begin{figure}[ht!]
  \centering
  \includegraphics[width=0.75\textwidth]{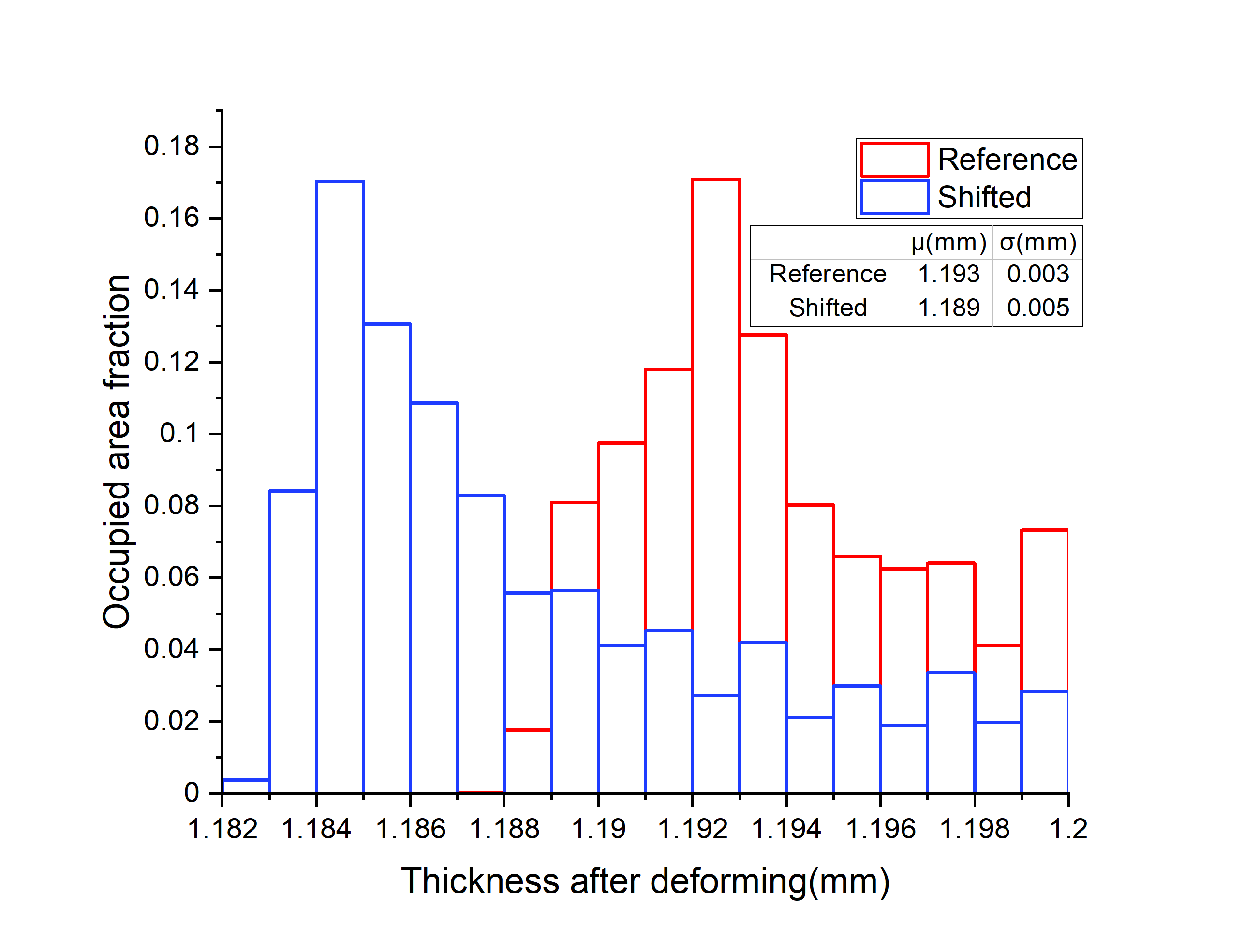}
  \caption{Distribution of the thickness of gas gap}
  \label{fig:def_distribution}
\end{figure}

The simulation of chamber deformation on the electrodes is carried out by using the pressure of the gas flow and an electric field between the two electrodes created by a \SI{6.6}{\kilo\volt} (working voltage of our RPC) potential difference. The deformation distributions of gas gap between two inner surfaces of electrodes for  the "Reference spacers" design and for the "Shifted spacers" design are shown in Figure~\ref{fig:deformation}.  And the distribution of the thickness of the gas gap after deforming for the "Reference spacers" and the "Shifted spacers" designs are shown in Figure~\ref{fig:def_distribution}.

\vspace*{0.3cm}

\begin{figure}[htp!]
  \centering
  \begin{subfigure}{.47\textwidth}
    \centering
	\includegraphics[width=1\textwidth]{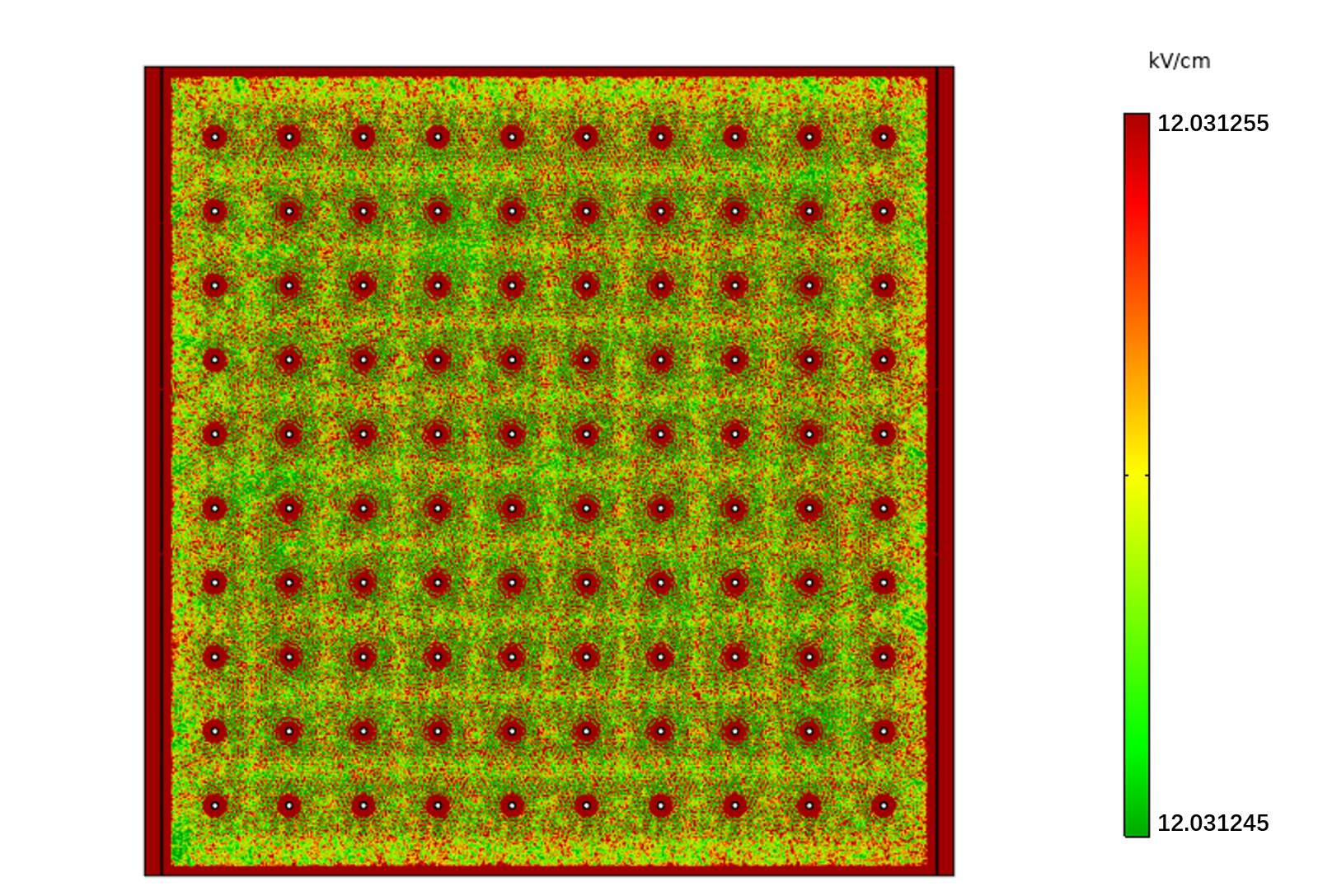}
	\caption{"Reference spacers" RPC.}
  \end{subfigure}
  \begin{subfigure}{.47\textwidth}
    \centering
	\includegraphics[width=1\textwidth]{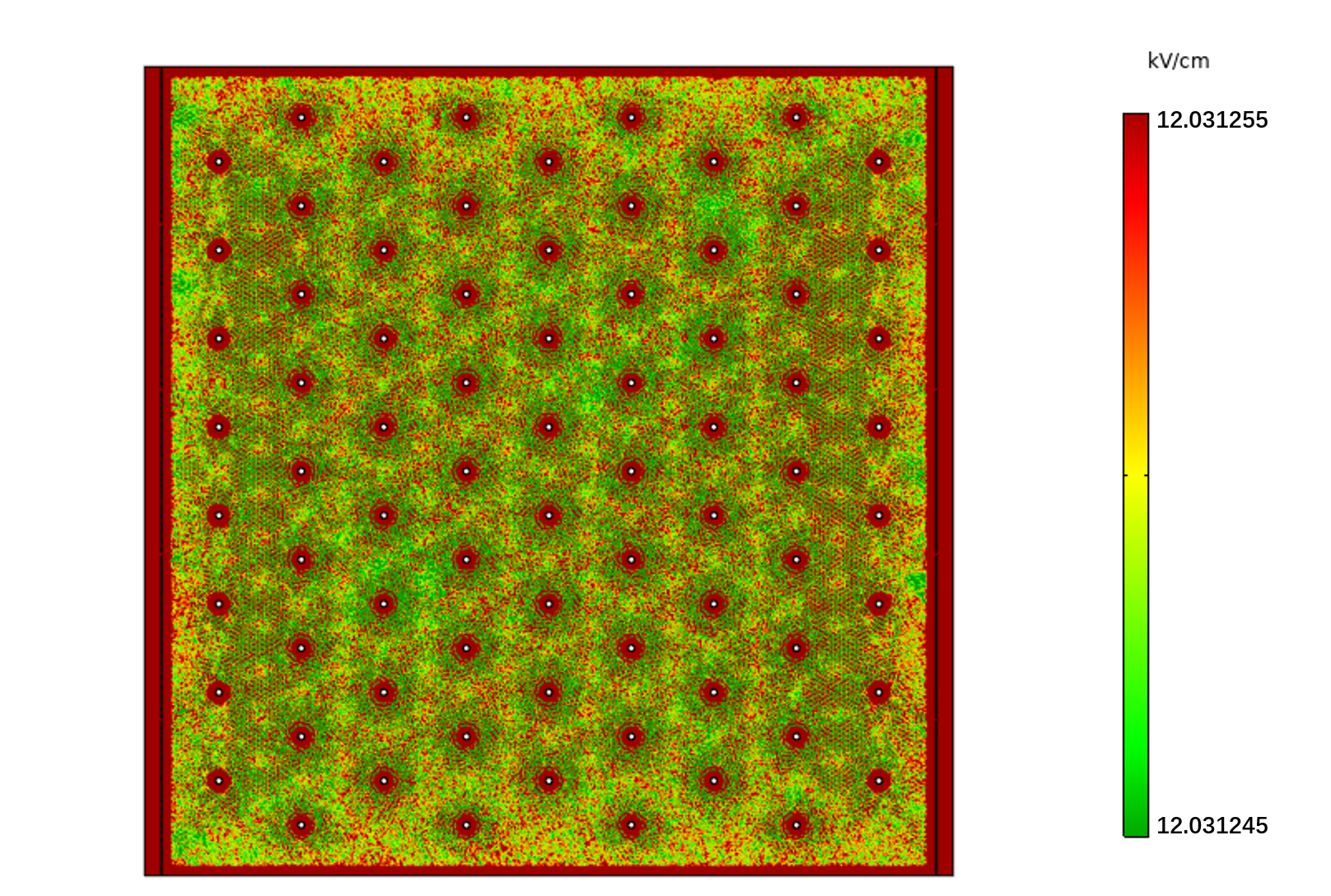}
	\caption{"Shifted spacers" RPC.}
  \end{subfigure}
  \caption{Electric field distribution inside the gas chamber.}
  \label{fig:electric_field}
\end{figure}

\newpage
The distributions of deformation are also calculated in the dashed line area. The mean thickness and RMS of the gas gap (\SI{1.2}{\milli\meter}) after deforming for the "Reference spacers" and the "Shifted spacers" RPCs are $1.193 \pm 0.003$ \si{\milli\meter} and $1.189\pm 0.005$ \si{\milli\meter}, respectively. The deformation uniformity, defined as the ratio of RMS and mean of deformation for "Reference spacers" and "Shifted spacers" RPCs, are 0.25\% and 0.42\%, respectively. The increase of the distance between spacers would cause more deformation on both electrodes, but the ratio of mean deformation and gas gap width is still within 1\%. Although the configuration of "Shifting spacers" decreases about 24\% of spacers, it still maintains similar deformation uniformity compared to "Reference spacers" RPC. The electric field distribution  inside the chamber after deforming is shown in Figure~\ref{fig:electric_field}.

\begin{table}[ht!]
  \centering
  \resizebox{1.0\textwidth}{!}
  {%
    \begin{tabular}{|c| c c |}
    \hline
	Model & “Reference spacers” RPC & “Shifted spacers” RPC \\
	\hline
	Mean velocity $\bar{v}$ & 0.238 (\SI{}{\milli\meter\per\second}) & 0.241 (\SI{}{\milli\meter\per\second}) \\
	RMS of velocity $\sigma_v$ & 0.049 (\SI{}{\milli\meter\per\second}) & 0.042 (\SI{}{\milli\meter\per\second}) \\
	$\sigma_v/\bar{v}$ & 20.3~(\%) & 17.5~(\%) \\ \hline
	Mean vorticity near spacers region & 0.0199 (\SI{}{\per\second}) & 0.0196 (\SI{}{\per\second}) \\
    RMS of vorticity near spacers region & 0.0129 (\SI{}{\per\second}) & 0.0127 (\SI{}{\per\second}) \\
	Mean vorticity excluding the vicinity of spacers & 0.0022 (\SI{}{\per\second}) & 0.0018 (\SI{}{\per\second}) \\  
    RMS of vorticity excluding the vicinity of spacers & 0.0028 (\SI{}{\per\second}) & 0.0026 (\SI{}{\per\second}) \\ \hline
	Mean thickness between gas gap  $\bar{d}$ & 1.193 (\SI{}{\milli\meter}) & 1.189 (\SI{}{\milli\meter}) \\
	RMS of deformation  $\sigma_d$ & 0.003 (\SI{}{\milli\meter}) & 0.005 (\SI{}{\milli\meter}) \\
	$\sigma_d/\bar{d}$ & 0.25~(\%) & 0.42~(\%) \\
	\hline
	\end{tabular}%
  }%
  \caption{\label{tab:simulation} Results from simulation.}
\end{table}

The summary of simulation results are listed in table \ref{tab:simulation}.  As discussed, while the "Shifted spacers" design decreases the number of spacers inside the chamber from \num{100} to \num{76},  by about $\sim\SI{24}{\percent}$, it can also provide better homogeneity for the gas velocity and lower vorticity.

\section{RPC Construction}
\begin{figure}[!ht]
  \centering
  \begin{subfigure}{.48\textwidth}
    \centering
	\includegraphics[width=0.86\textwidth]{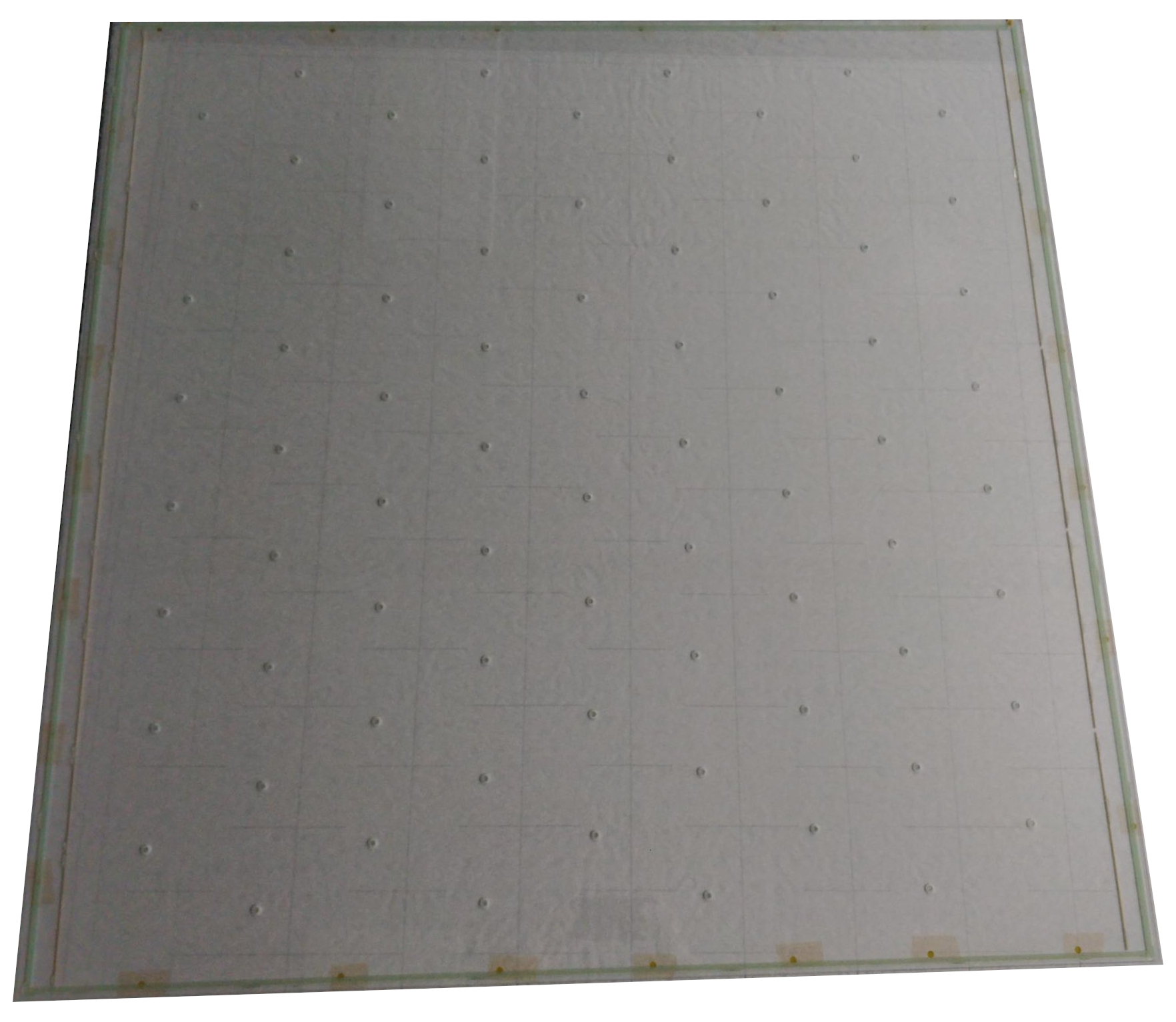}
  \end{subfigure}
  \begin{subfigure}{.48\textwidth}
    \centering
	\includegraphics[width=1\textwidth]{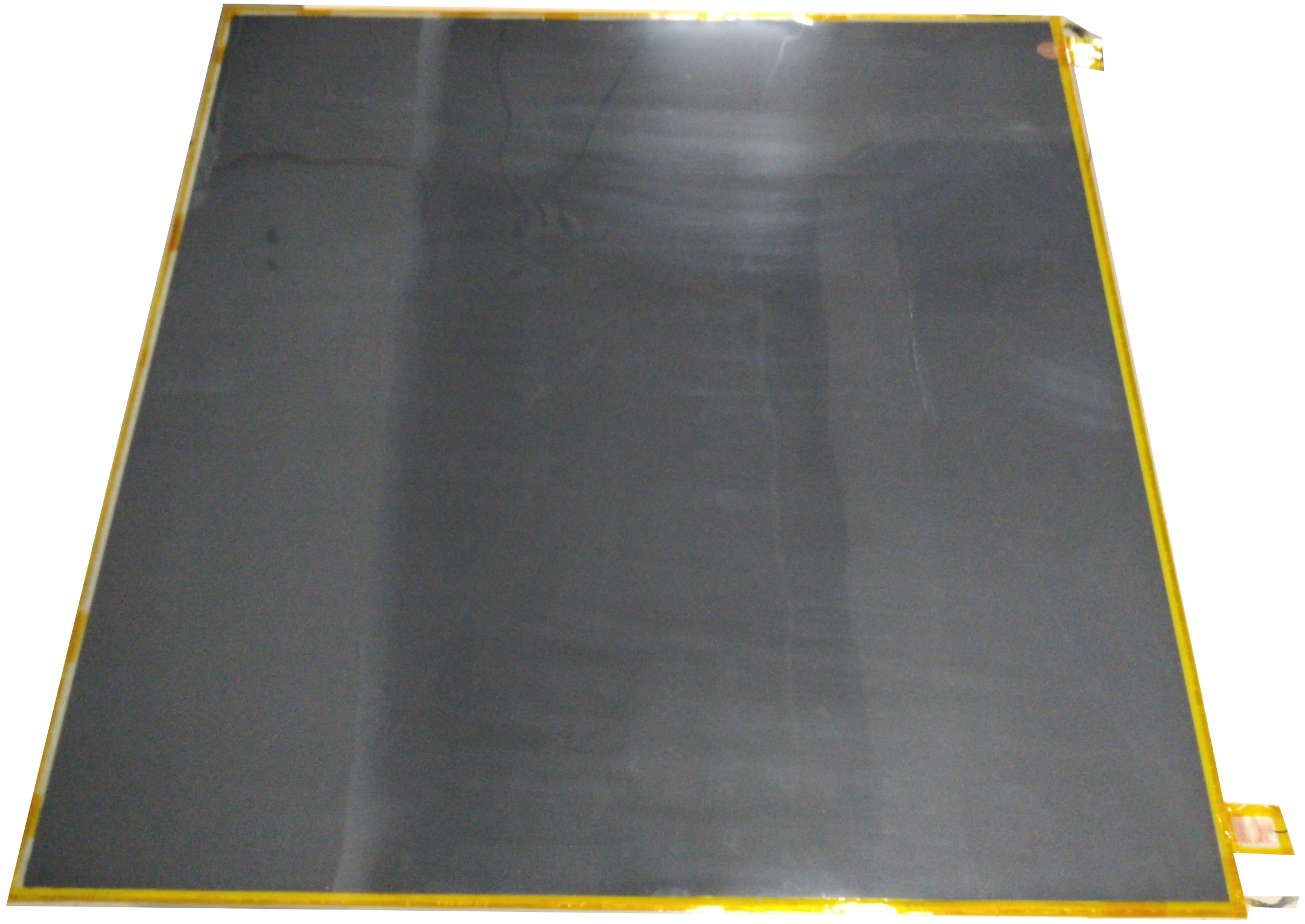}
  \end{subfigure}
  \caption{Construction of "Shifted spacers" Glass RPC.}
  \label{fig:construction}
  \vspace{-0.4cm}
\end{figure}

Based on the simulation results, both "Reference" and "Shifted" RPCs have been built in the clean room with size of $\num{1}\times\SI{1}{\square\meter}$. The construction procedure (Figure~\ref{fig:construction}) is listed as follows :

\begin{itemize}
  \item Glue the spacers, walls and PEEK bars on the \SI{1.1}{\milli\meter} electrode.
  \item Glue the spacers and the walls on the \SI{0.7}{\milli\meter} electrode and add silicone along the chamber walls.
  \item Paint glass with a graphite coating (a mixture of conductive graphite coating diluted with a lacquer) using a spray gun.
  \item Cover the graphite coated glass with the \SI{200}{\micro\meter} mylar films.
  \item Stick the power connectors on both sides of the chamber and tape the mylar films.
\end{itemize}

\vspace{-0.4cm}
\section{Performance Test of the RPCs}

\subsection{Cosmic muon test platform}

\begin{wrapfigure}[13]{r}{0.45\textwidth}
  \centering
  \includegraphics[width=0.48\textwidth]{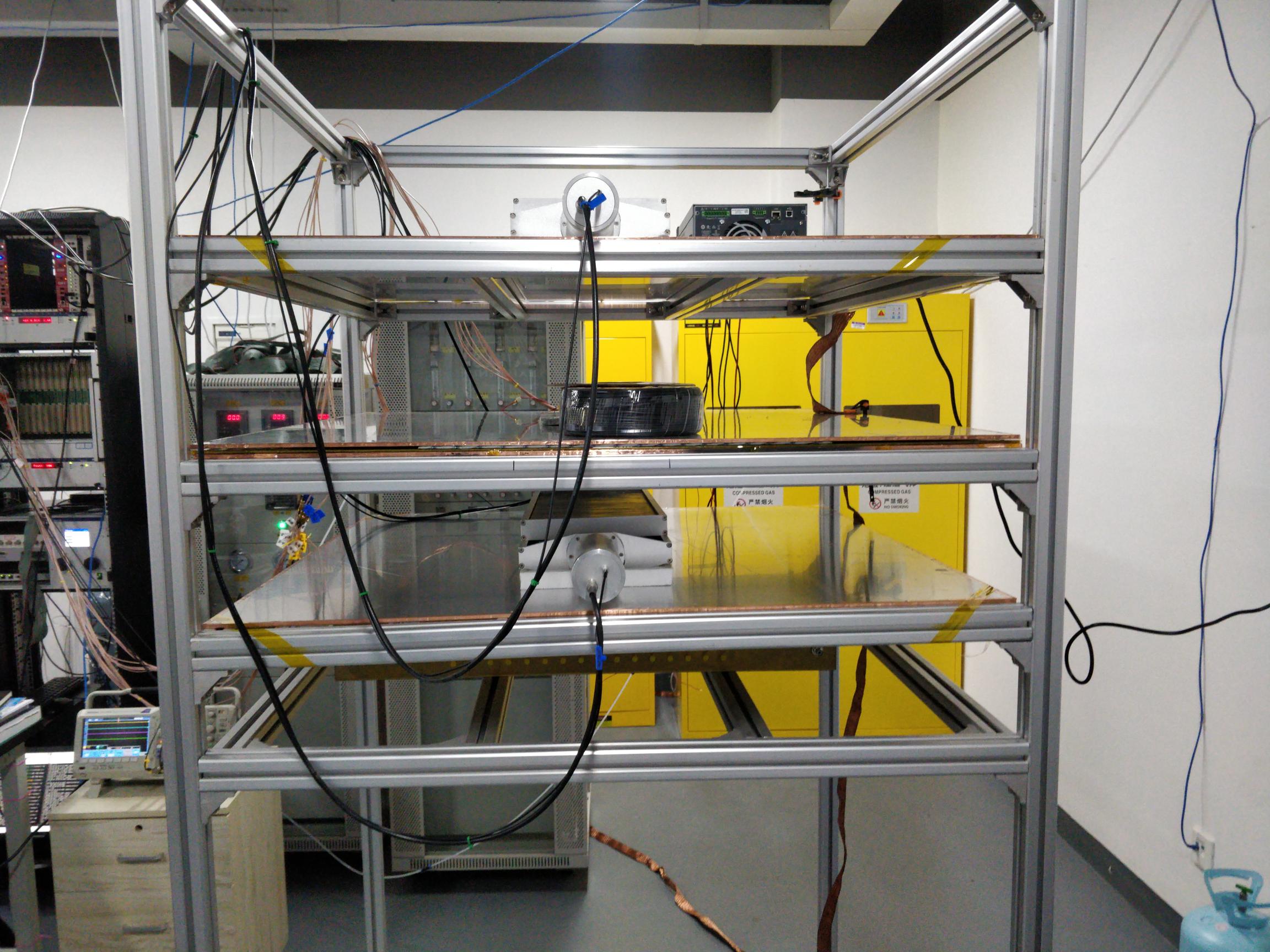}
  \caption{Cosmic test system}
  \label{fig:setup}
\end{wrapfigure}%
\sisetup{per-mode = symbol}%

The glass RPC signals induced by cosmic muons are collected on the cathode electrode with a \SI{3}{\milli\meter} thick foam panel covered with \SI{2.5}{\centi\meter} width copper strips. 
The signal from one side of each strip is amplified by a common emitter amplifier Front End Board (\textsf{FEB}) with a gain of $\sim$\num{16}, and then digitized by a \textsf{CAEN V172} \cite{caen} at \SI{5}{GS\per\second}. The other side of each strip is connected to the ground by a \SI{20}{\ohm} resistance to match the impedance of the strip panel and avoid reflections.

The triggering system is composed of four \num{40}$\times$\SI{20}{\square\centi\meter} scintillators. We place two scintillators on both side of the RPC chamber for signal coincidence (Figure~\ref{fig:setup}).

The scintillator trigger system is fully covered by the area of eight strips for the signal readout of RPC.

\begin{figure}[ht!]
  \centering
  \includegraphics[width=1\textwidth]{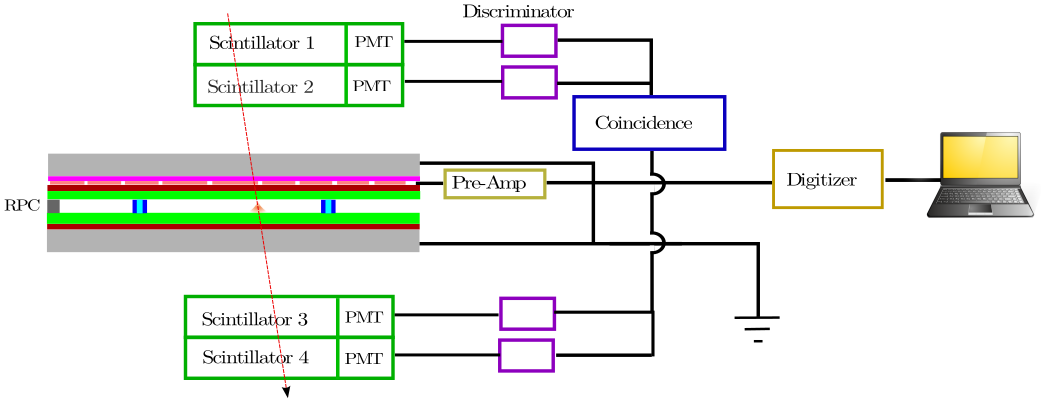}
  \caption{Schematic of the cosmic test platform (the coincidence unit triggers the digitizer for readout).}
  \label{fig:schematic}
\end{figure}

The RPC chamber is filled with a gas mixture, which includes \num{94.7}\% \ch{CH2FCF3} (R134a), \num{5}\% \ch{C4H10} and \num{0.3}\% \ch{SF6}. We use a gas flow rate corresponding to about one gas volume replacement every hour. The high voltage (HV) is provided by an \textsf{EHS 8080n} from \textsf{iseg} \cite{iseg}. A schematic of the cosmic test platform is shown in Figure~\ref{fig:schematic}.

\vspace*{-0.3cm}
\subsection{Data taking and analysis}

\begin{figure}[!ht]
  \centering
  \includegraphics[width=1.\textwidth]{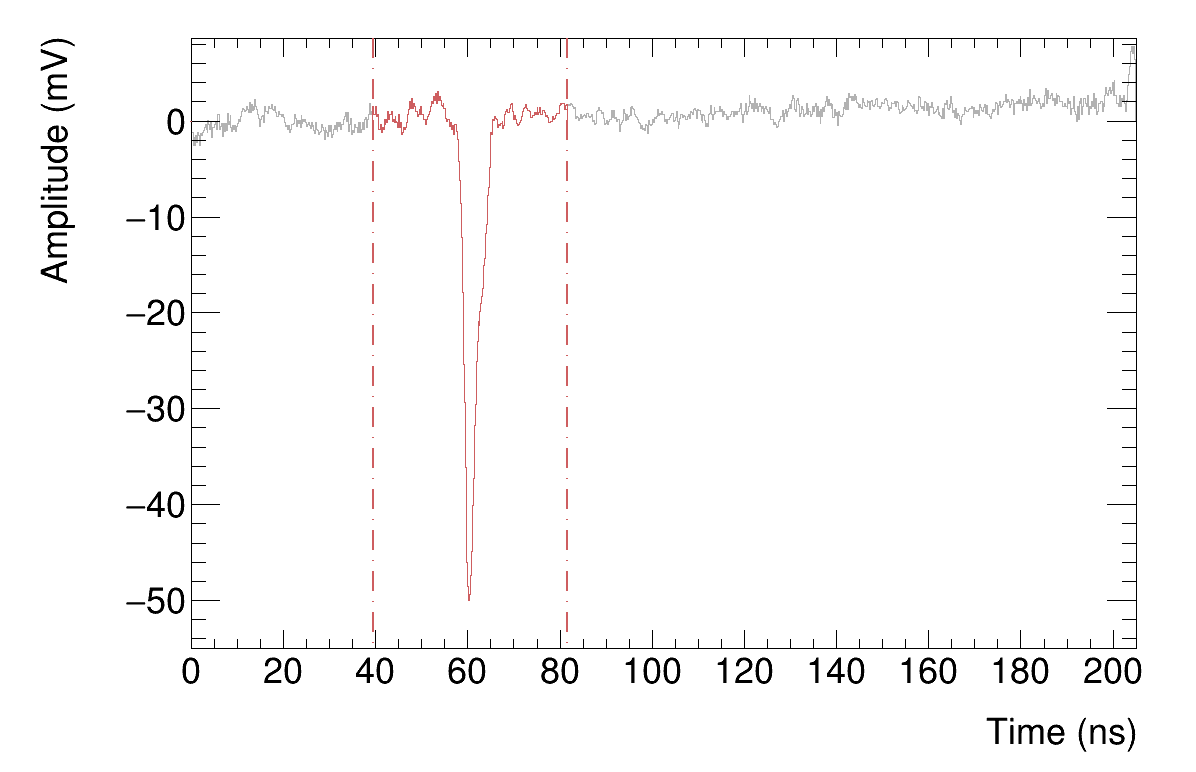}
  \caption{Event with a typical RPC signal (red curve).}
  \label{fig:signal}
  \vspace*{0.5cm}
\end{figure}

We studied the dependence of the RPC detection efficiency on the applied high voltage ranging from \SI{5700}{\volt} to \SI{7000}{\volt}, with a step of \SI{100}{\volt}. All scintillator-triggered cosmic muon events are recorded. Each event contains a digitized output of all eight readout strips and the scintillator trigger. The muon detection efficiency ($\epsilon$) of RPC can be calculated as follows :

\begin{equation}
 \epsilon = \frac{N_{signal}}{N_{triggered}},
\end{equation}

where $N_{signal}$ is the number of events with RPC signal, $N_{triggered}$ is the total number of triggered events by scintillators. Figure~\ref{fig:signal} shows an event with one signal (red curve). For each event, the $\SI{20}{\nano\second}$ region before the $\SI{40}{\nano\second}$ wide signal region was used to calculate the average noise $B$ and RMS $\sigma_{B}$ ($\sim\SI{0.9}{\milli\volt}$).  

\begin{figure}[!ht]
  \includegraphics[width=\textwidth]{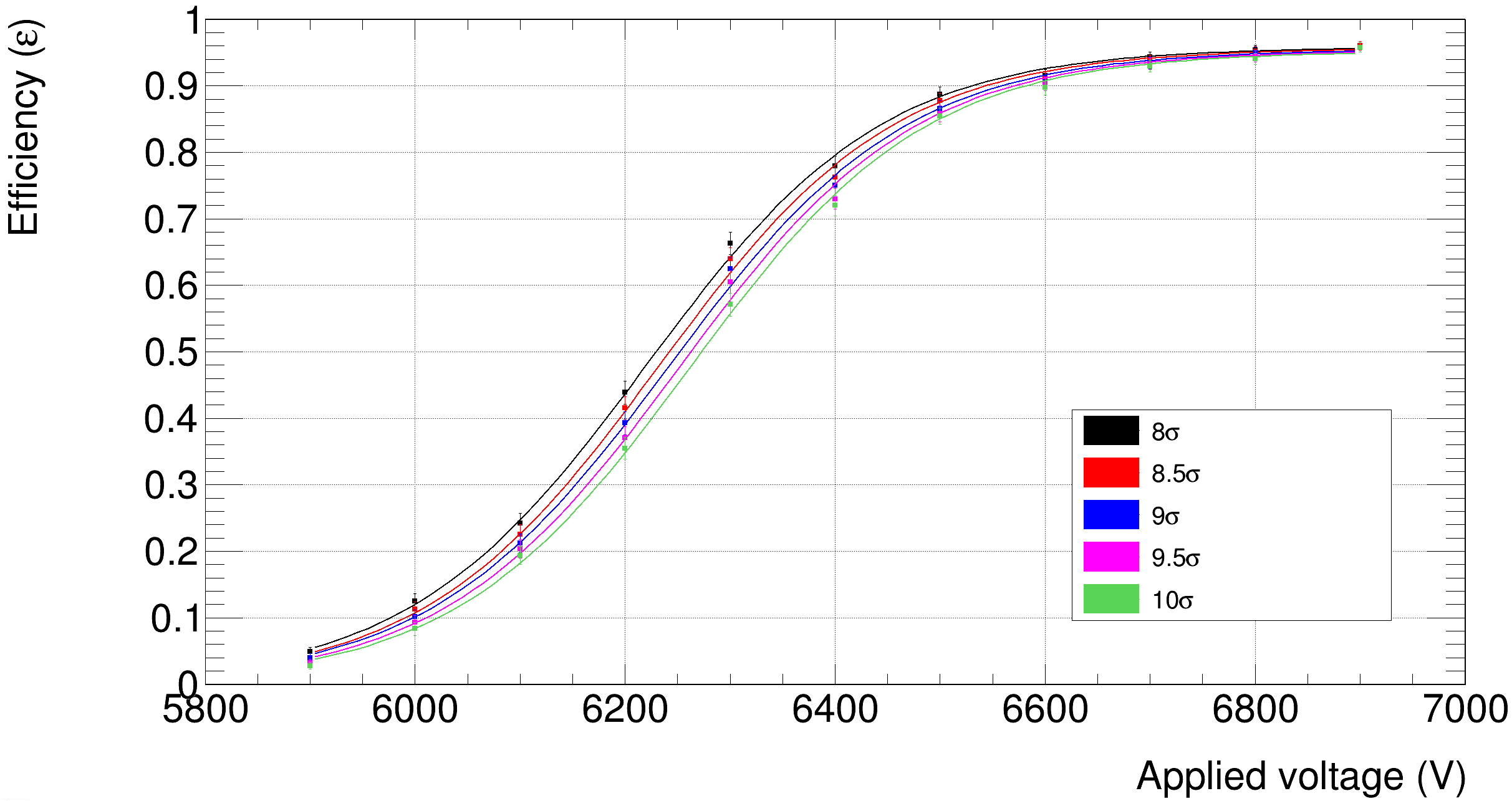}
  \caption{Efficiency versus applied high voltage for the "Reference spacers" chamber.}
  \label{fig:efficient}
    \vspace*{0.5cm}
\end{figure}

The chamber is considered efficient for this event if at least one channel has an amplitude greater than $N\times\sigma_{B}$. We performed the analysis for different values of $N$. The corresponding efficiencies are shown in Figure~\ref{fig:efficient}. They present consistent results after reaching the efficiency plateau for HV greater than 6800 Volt, which demonstrates the robustness of the muon detection against noise.  In this paper we use the value $N=8$.

The multiplicity is defined as the mean number of strips fired when the chamber has detected signal. Our study shows the multiplicity is around 1.3 for HV at 6800 Volt.


\newpage
\section{Results}

\begin{figure}[!ht]
  \centering
  \includegraphics[width=1\textwidth]{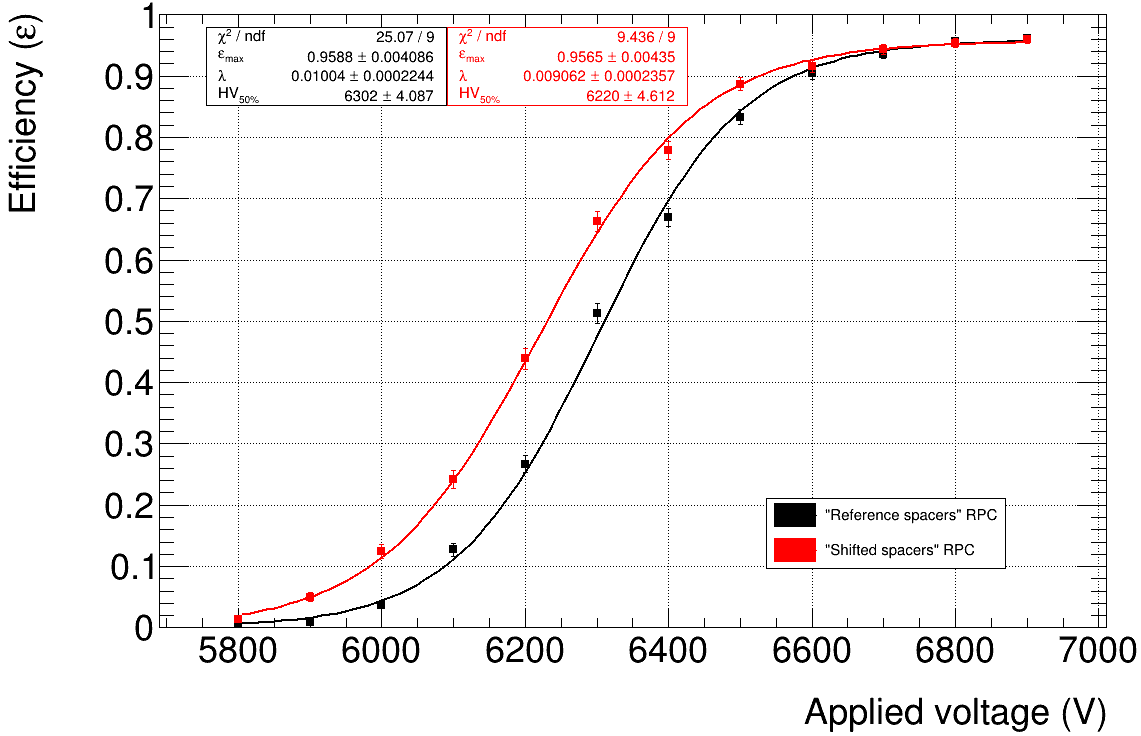}
  \caption{Efficiency versus applied voltage for "Reference spacers" and "Shifted spacers" chambers.}
  \label{fig:Plot}
  \vspace*{-0.2cm}
\end{figure}

The efficiencies with respect to the applied high voltage are shown in Figure~\ref{fig:Plot}. The "Shifted spacer" chamber is compared with the "Reference spacers"  chamber. They show very similar efficiency curves. More importantly, the efficiency plateaus are comparable for the two chambers, both greater than 95\%. No corrections for temperature and atmospheric pressure variations have been made, which largely explains the shift observed. It means the new design of RPC chamber maintains an equally good muon detection performance even with significantly less spacers used.


\section{Summary and Conclusions}

Optimization of the RPC gas chamber design has been performed, by using simulations of "Reference spacers" and "Shifted spacers" chambers. The simulations show that an optimized configuration of spacers can decrease the number of spacers by $~24\%$ while improving the gas flow and maintaining similar uniformity of the electrodes deformation. The tests performed on a cosmic test platform show that the "Shifted spacers" RPC maintains a high muon detection efficiency ($>95\%$) at high voltage, while using 24\% less spacers than the "Reference spacers" one.

\newpage
\acknowledgments
This study was supported by National Key Programme for S\&T Research and Development~(Grant NO.\ \ : 2016YFA0400400, 2016YFA0400100).

\normalem
\bibliography{bibli}
\end{document}